\input amstex
\documentstyle{amsppt}
\magnification=\magstep 1
\hsize29pc
\vsize42pc
\baselineskip=24truept

\def\pin{p_{\infty} (\lambda)}
\def\mi{\mu_{\infty}(\lambda)}

\def\psib{\overline{\psi}}
\topmatter
\title  A CONVEXITY THEOREM IN THE SCATTERING THEORY FOR THE DIRAC 
OPERATOR\endtitle
\author K.L.  Vaninsky\endauthor
\affil Kansas State University, Manhattan, KS 66506\endaffil
\email vaninsky@math.ksu.edu\endemail
\thanks The author would like to thank IHES where the paper was completed for hospitality.
The work is partially supported by NSF grant DMS-9501002\endthanks
\abstract The Dirac operator enters into zero curvature representation for the 
cubic nonlinear  Schr\"{o}dinger equation. We introduce and study a 
conformal map from the upper half-plane 
of the spectral parameter of the Dirac operator into itself. 
The action variables turn out to be limiting boundary values of the imaginary 
part of this map. We describe the image of the momentum map 
(convexity theorem) in the 
simplest case of a potential from the Schwartz class. 
We apply this description to the  invariant manifolds 
for the nonlinear Schr\"{o}dinger equation.
\endabstract
\endtopmatter
\rightheadtext{A CONVEXITY THEOREM IN THE SCATTERING THEORY }
\document

\subhead 1. Introduction \endsubhead The nonlinear Schr\"{o}dinger 
equation  (NLS) with 
repulsive potential\footnote"*"{$^\bullet$ denotes a
derivative in the time variable, $'$  in the $x$ variable.}
$$
i\psi^{\bullet}= -\psi'' + 2 |\psi |^2 \psi,
$$
has been solved for  square integrable  initial data on the circle 
and for  a sumable potential on the  entire line, see {\it e.g.} 
\cite{MCV, FT}. It is an 
important problem  which comes from  statistical mechanics \cite{MCV} 
to prove 
 complete integrability for   extensive class of initial data 
which accommodate some translation--
invariant Gibbs' distribution on the  entire line. 

The integration of the NLS equation on the entire line is based on the 
scattering theory for the Dirac operator --- the simplest $2\times 2$ canonical 
system. In this paper 
we introduce a special conformal map (CM) from the upper-half plane of 
the spectral parameter  into itself. The action variables,  known before for  
the rapidly decreasing 
case in terms of so-called scattering coefficients, are defined  now  
as  a boundary values of the imaginary part of this map.
The limiting
boundary values are  not a problem: they   are  known always
to exist for a positive harmonic function \cite{KO}.
What we propose, in fact, is a continuum analog of the
``Marchenko-Ostrovskii'' conformal map \cite {MO}.

It is known that the Dirac operator is self-adjoint and that the 
Weyl-\break Titchmarsh 
functions are unique under  sole condition of continuity of the potential.
We establish a relation between  the boundary values of the CM with the 
Weyl-Titchmarsh functions, with the hope that, suitably interpreted, 
this relation 
defines a CM for  sufficiently  broad class of potentials. 

Construction of actions for the NLS equation with  
initial data from a particular function 
class reduces  to the description of all possible boundary values of the CM, 
{\it i.e.} to the image of the momentum map. In this paper, we  prove  
the simplest theorem 
of that type:  for the Schwartz' class $S(R^1)$ of rapidly decreasing 
infinitely smooth complex functions, the image of the momentum map is 
a convex cone of real non-negative functions likewise from $S(R^1)$. This is an 
infinite--dimensional generalization  of the convexity theorem of 
Atiyah-Guillemin-Sternberg \cite{AT, GS}. As an application of our result 
we solve the following problem. 
\proclaim{\smc Problem}   What values of conventional local integrals of motion 
$H_1, H_2, \cdots,$ \break $ H_N,\; N <\infty$, where
$$\align
2H_1 &= \int |\psi|^2\, dx,\\
4H_2 &= - i\, \int \overline{\psi} \psi' \, dx,\\
8H_3 &= \int |\psi'|^2 + |\psi |^4\, dx    \quad  \text{etc. }
\endalign
$$
correspond some invariant manifold in $S(R^1)$ ? 
What are  all the invariant  
manifolds in $S(R^1)$ with given values of  $H_1, H_2, \cdots, H_N$. 
\endproclaim

The paper is organized as follows. In sections 2 and 3 we present
standard facts concerning Hamiltonian formalism and scattering theory 
needed in our considerations. 
The conformal map itself is introduced in section 4. 
It turns out conventional integrals of NLS  are the 
moments of the positive measure defined by  the imaginary part of the 
CM.  Section 5 establishes a  relation between the boundary values 
of the CM,   Weyl-Titchmarsh functions and the spectral matrix. 
Section 6 contains necessary information about the periodic problem 
for the NLS equation. The relation between the CM and "Marchenko-Ostrovskii 
picture" 
is presented in  section 7. We  also consider  
the whole structure in the linear limit. An interpretation of the results 
from the point of view of scattering theory is given in  section 8. 
In the  two sections 9 and 10, 
we show that the actions are canonically paired to the angles, which are 
given in terms of the phase of the reflection coefficient. 
This demonstrates an action of the torus with continuum number of generators,
a standard requirement in the convexity theorems.
We exploit
this structure to give an effective  solution of the moment problem.
The convexity theorem is presented in  section  11. A conjecture about
the image of the momentum map for potentials from the Sobolev'   class 
$H^s,\,  s\geq 0$ is also stated there. The last section 12 contains the 
solution of the problem. 

The author is grateful to P. Deift, L. Faybusovich, H. McKean,  V. Peller and 
N. Young for discussions.

\subhead 2. Hamiltonian formalism and zero curvature representation \endsubhead
\footnote"*"{The commutator representation for NLS equation was found by 
\cite{ZS}; see also  \cite{AKNS} for a more systematic approach.}
We denote by $S(R^1)$ the Schwartz' space of complex rapidly decreasing 
infinitely differentiable functions on the  line such that 
$$
\sup_{x} |(1+x^2)^n \, f^{(m)}(x)|\, < \infty\quad \text{for  all }\quad 
n,m=0,1,\cdots.
$$
On the phase space $M \equiv S(R^1)\times S(R^1)$, which is a 
product of  two copies of Schwartz' space,  we introduce a Poisson bracket
$$
\{A,\,B\}= \, i\, \int_{R^1} {\partial A\over \partial \psi}
{\partial B\over \partial \overline{\psi}}- {\partial A\over \partial \overline{\psi}}
{\partial B\over \partial \psi} dx.
$$
The bracket is well-defined for any functionals $A, B$ with  ``nice'' 
gradients. The original 
NLS equation 
$$
i\psi^{\bullet}= -\psi'' + 2 |\psi |^2 \psi,
$$
can be written in the form 
$$
\psi^{\bullet}= \{8 H_3,\, \psi\} \quad \text{with} \quad 8 H_3= \int_{R^1} 
|\psi'|^2 + |\psi |^4 dx = \text{\it energy}.
$$
There are two other classical integrals $2 H_1=\int |\psi |^2 dx =${\it number 
of particles} and $4 H_2= - i\, \int \overline{\psi} \psi' dx= ${\it momentum}. 
$H_1,\, H_2$ and $H_3$ are the first three   
Hamiltonians of an infinite hierarchy of commuting flows.
NLS is a consistency condition for  the zero-curvature representation 
$$
{\partial U \over \partial t}- {\partial V_3\over \partial x} + [U,\, V]= 0,
$$
where
$
U= U_0+ \lambda U_1
$ with 
$$
U_0=\left(\matrix
0 & \overline{\psi}\\
\psi & 0 
\endmatrix \right) ,\quad U_1= -i \sigma_3
$$
and 
$V_3=V_0+\lambda V_1+\lambda^2 V_2$ with
$$V_0=i\left(\matrix
|\psi|^2 & - \overline{\psi}'\\
\psi' & -|\psi|^2
\endmatrix \right), \quad V_1=-2U_2, \quad V_2=-4U_1.
$$
Here and below  $\sigma$  denotes the {\it Pauli matrixes}
$$
\sigma_1=\left(\matrix
 0& 1\\
1 & 0
\endmatrix \right), \quad 
\sigma_2=\left(\matrix
0 & -i\\
i &  0
\endmatrix \right), \quad 
\sigma_3=\left(\matrix
1 & 0\\
0 & -1
\endmatrix \right).  
$$

\subhead 3. The reduced monodromy matrix \endsubhead
The results of this section are basically well known. They are included 
here to make the paper  self-contained. 
\footnote"*"{A study of the scattering theory for 
canonical systems goes back to \cite{K,KMA,MA}. For the latest results see 
\cite{BDZ}.}
The standard monodromy matrix  
$M(x, y, \lambda)$ is the $2\times 2$  solution of 
$$
M'(x,y,\lambda)= U(x,\lambda) \, M(x,y,\lambda), \quad  \quad 
M(x,y,\lambda)|_{x=y}=I,
$$
where $I$ is the identity matrix.
The reduced monodromy matrix $T(x,y,\lambda)$ is defined as  
$$
T(x,y,\lambda)\equiv  E^{-1}
(\lambda x) M(x,y,\lambda) E^{-1}(-\lambda y),
$$ 
where $E(\lambda x)= \exp{(-i\lambda x \sigma_3)}$ is a solution of 
the free equation ($\psi\equiv 0$).
The matrix $T$ solves
$$
T'(x,y,\lambda)=U_0(x)\, E(2\lambda x) \, T(x,y,\lambda),\quad  
\quad T(x,y,\lambda)|_{x=y}= I.
$$
It can be expressed as 
$$
T(x,y,\lambda)= \overset\curvearrowleft\to\exp
\, \int_y^x U_0(\xi) E(2\lambda\xi) d\xi.
$$
Note that both $M$ and $T$ belong to  $SL(2,{\Bbb C})$.
The symmetry of the matrix \break 
$U:\, \sigma_1 U_0(x)\sigma_1= \overline{U_0(x)}$ 
is inherited by 
$T:\; \sigma_1T(x,y,\overline{\lambda})\sigma_1=
\overline{T(x,y,\lambda)}$.

For   $\psi \in S(R^1)$ and $\lambda$ {\it real}, 
the existence of the limit
$$
\lim T(x,y,\lambda)= T(\lambda)= \left(\matrix
a(\lambda) & b^{\star}(\lambda)\\
b(\lambda) & a^{\star}(\lambda) \endmatrix \right), \quad 
\text{for} \quad y \rightarrow -\infty \quad\text{and}\quad x 
\rightarrow +\infty
$$
follows easily from the inequality 
$$
| \overset\curvearrowleft\to\exp
\, \int_y^x U_0(\xi) E(2\lambda\xi) d\xi - I \, | \leq 
\exp\, \int_y^x | U_0(\xi) |  d\xi -1,
$$
where $|\bullet|$ is any  multiplicative matrix norm. 
Due to the symmetry of $T(x,y,\lambda),$\break   $T(\lambda)$   has the  form 
$$
\left(\matrix
a(\lambda) & \overline{b}(\lambda)\\
b(\lambda) & \overline{a}(\lambda)
\endmatrix \right) \quad \text{with} \quad |a(\lambda)|^2- |b(\lambda)|^2=1.
$$ 
The important  fact   needed bellow is that $ b(\lambda) \in S(R^1)$.
This implies the following property of $a(\lambda)$
\roster
\item"{\it i.}" For  $\lambda$ real  $|a(\lambda)| \geq 1$ and
                $|a(\lambda)|^2-1 \in S(R^1)$.
\endroster

To establish  properties of   $T(\lambda)$ for {\it complex} $\lambda$, we  
introduce   the {\it Jost solutions} $M_{\pm}(x,\lambda):$
$$
M_{\pm}'(x,\lambda)=U(x,\lambda)M_{\pm}(x,\lambda),\quad M_{\pm}(x)= 
E(\lambda x) + o(1)\quad \text{as}\quad x\rightarrow \pm \infty.
$$
They can be expressed through the solution of the free equation 
by means of  transformation operators:
$$
M_+(x,\lambda)=E(x\lambda)+ \int_x^{+\infty} \Gamma_+(x,\xi) 
E(\lambda \xi) d\xi,
$$
$$
M_-(x,\lambda)=E(x\lambda)+ \int_{-\infty}^x \Gamma_-(x,\xi) 
E(\lambda \xi) d\xi.
$$
Integral equations and estimates for the kernels 
$\Gamma_{\pm}$ is $2\times2$ are derived  following \cite{M, section 3.1}.

The kernels $\Gamma_{\pm}$ are unique and infinitely smooth in both variables. They  have a symmetry 
$\overline{\Gamma}_{\pm}= \sigma_1 \Gamma_{\pm} \sigma_1$ and admit  unique 
decompositions $\Gamma_{\pm}= C_{\pm} + A_{\pm}$, where $ C_{\pm}$, resp. 
$A_{\pm}$ commute, resp. anti-commute with $\sigma_1$. The components 
$C_-,\; A_-$ of the 
kernel $\Gamma_-$  satisfy 
$$\align
C_-(x,t)&= \int\limits_{-\infty}^{x} U_0(\xi) A_-(\xi,t+\xi -x)\, d\,\xi,\\
A_-(x,t)&= {1\over 2} U_0({t+x\over 2}) + \int\limits_{{t+x\over 2}}^{x} 
U_0(\xi) C_-(\xi, t+x-\xi)\, d\,\xi, \endalign
$$
whence
$$A_-(x,x)= {1\over 2} U_0(x), \quad \quad C_-(x,x)={1\over 2} 
\int\limits_{-\infty}^{x} U_0^2(\xi)\, d\,\xi.
$$
The components $C_+,\; A_+$ of the kernel $\Gamma_+$  also satisfy 
$$\align
C_+(x,t)&= -\int\limits^{+\infty}_{x} U_0(\xi) A_+(\xi,t+\xi -x)\, d\,\xi,\\
A_+(x,t)&= - {1\over 2} U_0({t+x\over 2}) - \int\limits^{{t+x\over 2}}_{x}
U_0(\xi) C_+(\xi, t+x-\xi)\, d\,\xi, \endalign
$$
whence
$$A_+(x,x)= -{1\over 2} U_0(x), \quad \quad C_+(x,x)={1\over 2}
\int\limits^{\infty}_{x}  U_0^2(\xi)\, d\,\xi.
$$

The monodromy matrix $M(x,y,\lambda)$ can be written in the form 
$$
M(x,y)=M_+(x)M^{-1}_{+}(y)= M_{-}(x)M^{-1}_{-}(y).
$$
Therefore
$$
M_{+}^{-1}(x)\, M(x,y)\, M_{-}(y)= M^{-1}_{+}(y)M_{-}(y)=M_{+}^{-1}(x)M_{-}(x).
$$
The expression at the right does not depend on $x$ or $y$ and it is equal to 
$T(\lambda)$ as one can see by passing to the limit  as 
$ x\rightarrow +\infty,\, y \rightarrow -\infty$.
Introducing the  notation $M_\pm=(M^1_\pm,\, M^2_\pm)$ for the columns, 
we get  
$a(\lambda)= \text{det}\,(M^1_-,\,M^2_+)$. Now the following properties of 
$a(\lambda)$  follow easily  from the integral representation of the 
Jost solutions
\roster
\item"{\it ii.}"  $a(\lambda)$ is analytic in the upper half-plane and
infinitely smooth up to the boundary;
\item"{\it iii.}" $a(\lambda) = 1 +o(1)$ as $ |\lambda| \longrightarrow \infty$
and  $a(\lambda)$ is  root-free.
\endroster

\subhead 4. The conformal map $p_\infty(\lambda)$ \endsubhead
Let $p_\infty(\lambda)$ be  such that $a(\lambda)= \exp(-ip_{\infty}(\lambda))$ 
for $\lambda$ in the upper half-plane. From the properties of $a(\lambda)$ 
presented above  follow  
\roster
\item"{\it i.}" $ \pin $  is analytic in the upper half-plane and 
infinitely smooth up to the boundary.
\item"{\it ii.}" $\Im \pin \geq 0$ for $ \Im \lambda \geq  0$.
\endroster
This means simply that  $\pin$
defines a conformal map from the upper 
half-plane of the spectral parameter into itself.
Moreover, 
\roster
\item"{\it iii.}" $\pin=o(1)$ for $|\lambda| \rightarrow \infty$; 
for real $\lambda$,  the 
density of the measure $d \mi= \Im \pin \, d \lambda$ belongs to  $S(R^1)$ and 
$d \mi$ has  moments of any order.
\endroster
This is due to the fact that $|a(\lambda)|^2-1 \in S(R^1)$. 
The function $\pin$ can be written in the form
$$
\pin={1\over \pi}\int {d \mu_{\infty}(t)\over t -\lambda},
$$
see \cite{A}.
Expanding the denominator in inverse powers of $\lambda$, we obtain:
$$
\pin= -\sum^{\infty}_{k=0}{1\over \lambda^{k+1}} {1\over \pi}\int_{-\infty}^{+
\infty} t^k d \mu_{\infty}(t)= -{H_1\over \lambda}- {H_2\over \lambda^{2}}-
{H_3\over  \lambda^3}-\dots .
$$
where $H_1,H_2$ and $H_3$ are the integrals introduced above. 
The expansion has an asymptotic character for $\lambda: \; \delta \leq 
\arg \lambda \leq \pi -  \delta$.
An identification of the coefficients of the asymptotic expansion 
employs  a recurrence relation for  $H$'s.  
As in  \cite{MCV}, a straightforward computation produces: 
$$
D{1\over 2\psib}\left[2i\lambda{\partial T\over \partial \psi (x)}+ D {\partial 
T\over \partial \psi(x)}\right]=\psi{\partial T\over \partial \psi}-
\psib{\partial T\over \partial \psib},
$$
$$
D{1\over 2\psi}\left[2i\lambda{\partial T\over \partial \psib (x)}- D {\partial 
T\over \partial \psib(x)}\right]=\psi{\partial T\over \partial \psi}-
\psib{\partial T\over \partial \psib}.
$$
Here we  have  used   only the formulas for derivatives:
$$
{\partial T\over \partial \psi(x)}=T(+\infty, x) \left(
\matrix
0 & 0\\
1 & 0
\endmatrix \right) E(2\lambda x) T(x,-\infty),
$$
$$
{\partial T\over \partial \psib(x)}=T(+\infty, x) \left(
\matrix
0 & 1\\
0 & 0
\endmatrix \right) E(2\lambda x) T(x,-\infty).
$$
Now  we can put the formulas into  desired form of a recurrence relation:
$$
J\nabla H_{n+1}= K\nabla H_{n}, \quad n=1,2,\dots , 
$$
with\footnote"*"{$D^{-1}=1/2\left(\int^x_{-\infty}-\int^{\infty}_{x}\right).$}  
$$
K=\left(\matrix
-i\psi D^{-1}\psi & -i/2D+i\psi D^{-1}\psib\\
-i/2D+i\psib D^{-1}\psi & -i\psib D^{-1}\psib   
\endmatrix \right) \quad \text{and}\quad  J=i \sigma_2.
$$
This allows us to reconstruct  the $H$'s starting from $2H_1=\int|\psi|^2$.  
The same recurrence relation implies an involutive character of  $H$'s; see
\cite{MC1}.

\subhead 5. The relation between $\pin$, Weyl-Titchmarsh functions and the 
spectral matrix \endsubhead 
By definition, the {\it Weyl solutions} 
$ h_{\pm}= \left( \matrix
h_{\pm}^1 \\
h_{\pm}^2 \endmatrix \right)
$ 
of $h'=Uh$ belong, resp., to
$L^2(R^1_\pm)$ for $\lambda$ with strictly positive imaginary part; 
they are  normalized by  
$h_+^2(x,\lambda)=1$, resp.,  $h_-^1(x,\lambda)=1$ for $x=0$. They can 
be written in the form
$$\align
h_+(x,\lambda) & = M^1(x,\lambda)m_+(\lambda) +M^2(x,\lambda), \\
h_-(x,\lambda) & = M^1(x,\lambda) +M^2(x,\lambda)m_-(\lambda),
\endalign$$
where $M(x,\lambda)\equiv M(x,0, \lambda)$. 
For example, for the free equation,  the {\it Weyl-Titchmarsh functions }
$m_\pm$ are identically zero and 
$$\align
h_+(x,\lambda) & =\left( \matrix 0\\
                    e^{i\lambda x}\endmatrix \right)\,  \in L^2(R^1_+),\\
h_-(x,\lambda) & =\left( \matrix e^{- i\lambda x}\\
                              0\endmatrix \right)\, \in L^2(R^1_-).
\endalign$$
It is well-know fact, see  \cite{LS}, that under the   sole assumption of the 
continuity of  the potential $\psi$, the functions $m_{\pm}$ are unique.
 
Our first goal is to prove that
$$
e^{2 \Im \pin}={|1-m_-m_+|^2\over (1-|m_+|^2)(1-|m_-|^2)} \quad 
\text{for }\,\, \lambda \,\, \text{real.}
$$
This formula provides an alternative way to define the CM.

Note that $M^2_+$ resp., $M^1_-$ are from $L^2(R^1_+)$ resp., $L^2(R^1_-)$ 
for $\lambda$ with 
strictly positive imaginary part. Therefore 
$$
M^1_-  =k_-(\lambda) h_- , \quad\quad 
M^2_+  =k_+(\lambda) h_+, 
$$
for some numbers $k_\pm(\lambda)$. Take  $\psi$  
supported in $[-L, +L]$. Then
$$
M(0,L)M_+^2(L)=M_+^2(0)=k_+h_-(0).
$$
In more detail,
$$
\left(\matrix
m_{11} & m_{12}\\
m_{21} & m_{22} \endmatrix \right) (0,L) 
\left(\matrix 
0\\
e^{i \lambda L} \endmatrix \right) =
\left(\matrix 
m_{+}\\
1 \endmatrix \right) k_{+},
$$
or 
$$
m_{22}e^{i\lambda L} = k_{+}, \quad m_{12} e^{i \lambda L} = m_{+} k_{+}.
$$
Finally, for $\lambda$ real
$$
1=|m_{22}|^2-|m_{12}|^2=|k_+|^2(1-|m_{+}|^2)\quad \text{and} \quad 
|k_+|^2={1\over 1-|m_+|^2}
$$
and likewise
$$
|k_-|^2={1\over 1-|m_-|^2}.
$$
Therefore, for such $\lambda$, 
$$\align
e^{2 \Im \pin }& = |a(\lambda)|^2=  | \text{det}\,(M^1_-,\,M^2_+)|^2=
|k_-k_+(1-m_-m_+)|^2 \\
&= {|1-m_-m_+|^2\over (1-|m_+|^2)(1-|m_-|^2)}, \endalign
$$
as required.

The first order system $F'=U F$ introduced at the beginning can be expressed 
as   an eigenvalue problem for the {\it Dirac operator} ${\frak D}$:
$$
{\frak D} f=\left[\left(\matrix 1 & 0 \\
                                0 & 1 \endmatrix \right) i D + 
             \left(\matrix 0  & -i \psib \\
                                i \psi  & 0 \endmatrix \right) \right] f =
\lambda f.
$$
\proclaim{\smc The Spectral Theorem} For the operator ${\frak D}$ with a 
continuous potential there exists a unique 
$2\times 2$ Hermitian spectral matrix 
$\|\rho_{ij}(\lambda)\|$; $\|\rho_{ij}(\lambda')-
\rho_{ij}(\lambda'')\|$ is positive definite for any real $\lambda' >
\lambda''$; the entries $\rho_{ij}(\lambda)$ 
are  functions continuous from the right and
$$
\int {|d \rho_{ij}(\lambda)|\over 1 +\lambda^2} < \infty.
$$
For any vector-function $f \in L^2(R^1)$, the sequence 
$$
F_n^i(\lambda)=\int^{+n}_{-n} f^T(x)\overline{M}^i(x,\lambda) dx,\quad i=1,2
$$
converges to some $F_i(\lambda)$ in $L^2(R^1, \rho_{ij})$:
$$
\int\sum_{ij}(F^i_n-F^i)\overline{(F^j_n- F^j)}d \rho_{ij} \rightarrow\infty, 
\quad \text{as } \quad n\rightarrow \infty.
$$
Moreover,
$$
\int |f|^2 dx= \int \sum_{ij}F^{i}(\lambda)\overline{F^j(\lambda)} d\rho_{ij}
(\lambda).
$$
The function $\rho_{ii},\quad i=1,2$ is the boundary value of the expression
$$
{1\over 2 \pi} {1- |m_-|^2|m_+|^2\over |1-m_-m_+|^2}
$$ 
and $\rho_{12}=\overline{\rho_{21}}$ is the boundary value  of the expression
$$
{1\over 2 \pi} {\overline{m_+}(1-|m_-|^2)-m_-(1-|m_+|^2)\over  |1-m_-m_+|^2}.
$$
\endproclaim
The proof is  by  standard arguments, see for example 
\cite{CL}. By  direct computation,
$$
\text{det}\|\rho_{ij}\|={1\over 4 \pi^2} {(1-|m_+|^2)(1-|m_-|^2)\over 
|1-m_-m_+|^2},
$$
which implies:\ 
 $\text{det}\|\rho_{ij}(\lambda)\|=  e^{-2 \Im \pin}/4 \pi^2$ for $\lambda $ 
real.\footnote"*"{H.P. McKean in \cite{MC2} proved a similar formula.}

One can think about  spectral matrix as a smooth functional on the phase 
space. Indeed, gradients of the functions $m_{\pm}$ can be computed, {\it e.g.} 
$$\align
{\partial m_+ (\lambda)\over \partial \psi(y)} &= \;\;\; \left[h_+^1(y,\lambda)\right]^2,\\
{\partial m_+ (\lambda)\over \partial \psib(y)} &= - \left[h_+^2(y,\lambda)\right]^2. \endalign
$$
We split the proof of  the first formula into 5 steps. 
\subsubhead Step 1 \endsubsubhead By direct computation 
$$ 
{\partial M(x,0)\over \partial \psi(y)}=M(x,y)\left(\matrix 0& 0\\
                                                            1& 0 \endmatrix \right) M(y,0),
$$
$$
{\partial M(x,0)\over \partial \psib(y)}=M(x,y)\left(\matrix 0& 1\\  
                                                             0& 0 \endmatrix 
\right) M(y,0),
$$
where $ 0 \leq y \leq x$.

\subsubhead Step 2 \endsubsubhead Let $f,g$ be  two vector solutions of the 
equation: ${\frak D} f=\lambda f$, 
then \break  
$\det [ f(x), g(x) ] $ does not depend on $x$. The proof is trivial. 

\subsubhead Step 3 \endsubsubhead  The Weyl solution $h_+$  normalized by:  
 $h^2_+(x,y,\lambda)|_{x=y} =1$, can be expressed 
$$
h_+(x,y,\lambda)= M^1(x,y,\lambda)m_+(\lambda,y) + M^2(x,y,\lambda),
$$
where $m_+(\lambda,y)$ is the corresponding   Weyl-Titchmarsh function. 
In particular,  \break
$h_+(x,\lambda)=h_+(x,0,\lambda)$. The fact needed below is that 
$$
m_+(\lambda,y)={h^1_+(y,0,\lambda)\over h^2_+(y,0,\lambda)}.
$$ 
To prove this, note that the identity $M(x,0)=M(x,y)M(y,0)$ implies 
$$\align
M^1(x,0)&= m_{11}(y,0) M^1(x,y) + m_{21}(y,0)M^2(x,y),\\
M^2(x,0)&= m_{12}(y,0) M^1(x,y) + m_{22}(y,0)M^2(x,y). \endalign
$$
Therefore, for $0\leq y \leq x$:
$$\align
h_+(x,0,\lambda)&= M^1(x,0)m_+(\lambda,0)+M^2(x,0)\\
                &=\left[ m_{11}(y,0) M^1(x,y) + m_{21}(y,0) M^2(x,y)\right]m_+(\lambda,0) \\
                &\quad\quad + \left[ m_{12}(y,0) M^1(x,y) + m_{22}(y,0) M^2(x,y)\right]\\
                &=\left[ m_{21}(y,0) m_+(\lambda,0) + m_{22}(y,0)\right] \\
                &\quad\quad \times  \left[M^1(x,y) {m_{11}(y,0) m_+(\lambda,0) + m_{12}(y,0)\over 
                m_{21}(y,0) m_+(\lambda,0) + m_{22}(y,0) } + M^2(x,y) \right]. \endalign
$$
The result follows. 
\subsubhead  Step 4 \endsubsubhead
The purpose of this  step is to prove the formula
$$
{\partial m_+(\lambda,0)\over \partial \psi(y)}=- m_+(\lambda,0)m_+(\lambda,y)\left[{m_{12}(y,0)\over 
A}+{m_{11}(y,0)\over B}\right],
$$
where 
$$\align
A&= \;\;\; m_{12}(y,0)-m_+(\lambda,y)m_{22}(y,0),\\
B&= -m_{11}(y,0)+ m_+(\lambda,y) m_{21} (y,0). \endalign
$$

Consider the eigenvalue problem: ${\frak D} f=\lambda f$ with 
the boundary conditions $f^2(y,\lambda)=1,\;\; f^1(b,\lambda)=f^2(b,\lambda)$, 
where $0\leq y < b < +\infty$. The solution can be written in the form 
$$
f(x,\lambda)= M^1(x,y,\lambda) m_+(\lambda,y,b) +M^2(x,y,\lambda).
$$
Due to the selfadjointness of the spectral problem, the limit 
$$
m_+(\lambda,y)=\lim\limits_{b\rightarrow +\infty} m_+(\lambda,y,b)
$$
exists and  does not depend on the boundary  condition at the right end of the interval. 
From the definition,
$$
m_+(\lambda,y,b) ={m_{22} -m_{12}\over m_{11}-m_{21}}(b,y,\lambda).
$$
Therefore,
$$\align
\nabla m_+(\lambda,0,b)&=\;\;\;{m_{22} -m_{12}\over m_{11}-m_{21}} 
                      {\nabla m_{22} -\nabla m_{12}\over m_{22}-m_{12}}
(b,0,\lambda)\\
 &\quad - {m_{22} -m_{12}\over m_{11}-m_{21}} 
                    {\nabla m_{11} -\nabla m_{21}\over m_{11}-m_{21}}(b,0,\lambda). \endalign
$$
By Step 1,
$$\align
{\partial m_+(\lambda,0,b)\over \partial \psi(y)} &= \;\;\;m_+(\lambda,0,b) 
                         {(m_{22} -m_{12})(b,y)m_{12}(y,0)\over (m_{22}-m_{12})(b,0)} \\
          &\quad - m_+(\lambda,0,b) {(m_{12} -m_{22})(b,y)m_{11}(y,0)\over (m_{11}-m_{21})(b,0)} \\
                     &=m_+(\lambda,0,b)(m_{22}-m_{12})(b,y)\\ 
             &\quad \times  \left[{m_{12}(y,0)\over (m_{22}-m_{12})(b,0)} 
              + {m_{11}(y,0)\over (m_{11}-m_{21})(b,0)} \right].
\endalign
$$
Using the identity $M(b,0)=M(b,y)M(y,0)$,  and simple algebra, one finds
$$\align
-m_{+}& (\lambda,0,b) m_{+}(\lambda,y,b)\\
 \times & \left[  {m_{12}(y,0)\over m_{12}(y,0) - m_+(\lambda,y,b)m_{22}(y,0) } +{m_{11}(y,0)\over 
-m_{11}(y,0) +m_+(\lambda,y,b)m_{21}(y,0)} \right]. \endalign
$$
Now pass to the limit $b\rightarrow \infty$.
\subsubhead Step 5 \endsubsubhead
By  Step 3,
$$
m_+(\lambda,y)[m_{21}(y,0)m_+(\lambda,0)+m_{22}(y,0)]= 
m_{11}(y,0)m_{+}(\lambda,0)+m_{12}(y,0).
$$
After simple algebra,
$$
{m_+(\lambda,0)\over A}={1\over B}.
$$
Therefore,
$$
{\partial m_+(\lambda,0)\over \partial \psi(y)}=-m_+(\lambda,y)\left[{m_{+}(\lambda,0) 
m_{11}(y,0) + m_{12}(y,0)\over -m_{11}(y,0) + m_{21} (y,0) m_+(\lambda,y)}\right].
$$
By Step 3,
$$\align
{\partial m_+(\lambda,0)\over \partial \psi(y)}=
&{h^1_+(y,\lambda)\over m_{11}(y,0)\dsize\frac{h^2_+(y,0)}{h^1_+(y,0)} - m_{21}(y,0)}\\
=&{ \left[h_+^1(y,\lambda)\right]^2\over m_{11}(y,0)h_{+}^2(y,0)-m_{21}(y,0)h^1_{+}(y,0)}. \endalign
$$
By Step 2, the denominator can be computed at $y=0$, where it is equal to 1. We are done.

The proof of the second formula is the same.

\subhead 6. The necessary information about periodic 
problem\footnote"*"{For more information  see \cite{MCV}.} \endsubhead
In the next two sections we relate the  CM defined by $\pin$ 
with  the analogous object known in the periodic case.
Consider the  periodic problem for the NLS, {\it i.e.}  $x$ is on the circle 
of  perimeter $2L$ and  assume that the function $\psi$ is 
infinitely smooth. 
The  {\it periodic/anti-periodic}, spectrum of the auxiliary 
spectral problem is determined 
by  $$\Delta_{L}(\lambda)\equiv \text{trace} M(L, -L, \lambda)/2.$$ 
Namely,  $\lambda_0$ is a {\it periodic}, resp. {\it anti-periodic}, 
eigenvalue if it is a root of the equation: 
$\Delta_L(\lambda_0)=+1$, resp. $-1$. The function $\pi_L(\lambda)$ is defined 
by: $\Delta_L(\lambda)=\cosh i \pi_L (\lambda)$; it has   the properties  
\roster
\item"{\it i.}" $\pi_L(\lambda)$ is single-valued on the $\lambda$-plane 
cut along open gaps ($\lambda \in R^1: \quad |\Delta_L(\lambda)| > 1$); 
\item"{\it ii.}" $\bar{\pi}_L(\lambda)=\pi_L(\bar{\lambda})$;
\item"{\it iii.}" $\Im \pi_L(\lambda) \geq 0$ for $\lambda$ in the upper 
half-plane. 
\endroster
The function $\pi_L(\lambda)$ determines a conformal map from  the upper 
half-plane into a 
slit-domain. In fact, this CM was known to physicists  already in  1929, see 
\cite{B}.
This map was used by \cite{MO} in their  description of  all possible 
spectral sequences for the periodic Schr\"{o}dinger operator.

The function $\pi_L(\lambda)$ can be written in the form 
$$
\pi_L(\lambda)= 2 \lambda L + p_L(\lambda)
$$
in which
\roster
\item"{\it i.}" $p_L(\lambda)$ is analytic and continuous up to the boundary;
\item"{\it ii.}" $\bar{p}_L(\lambda)=p_L(\bar{\lambda})$;
\item"{\it iii.}" $\Im p_L(\lambda) \geq 0$ for $\lambda$ in the upper 
half-plane; the inequality is strict, except for a simple zeroes, 
one in each band.
\endroster
Introducing a measure $d\mu_L(\lambda)\equiv \Im p_L(\lambda)d\lambda,$ we 
can write $p_L(\lambda)$ in terms of the Cauchy integral
$$
p_L(\lambda)={1\over \pi} \int{d\mu_L(t)\over t- \lambda}.
$$
The function $p_L(\lambda)$ has an asymptotic expansion at infinity 
analogous to the asymptotic expansion of $\pin$:
$$
p_L(\lambda)= \pi_L(\lambda)- 2 \lambda L = -\sum^{\infty}_{k=0}{1\over 
\lambda^{k+1}} {1\over \pi}\int_{-\infty}^{+
\infty} t^k d \mu_{L}(t)= -{H_1\over \lambda}- {H_2\over \lambda^{2}}-
{H_3\over  \lambda^3}-\dots .
$$

\subhead 7. The infinite-volume limit. Deformation to the linear problem \endsubhead
The scaling $\psi\rightarrow \sqrt{\varkappa}\psi, \;\,
 \varkappa >0$, transforms the original NLS equation into
$$
i\psi^{\bullet}= -\psi'' + 2 \varkappa |\psi |^2 \psi.
$$
We denote by $p^\varkappa_\infty(\lambda), \; \psi_L^\varkappa(\lambda),
$ {\it etc.,} all the 
quantities introduced before to emphasize their dependence on the parameter 
$\varkappa$. The function $ p^\varkappa_\infty(\lambda)$ turns out to be a 
continuum analog of $p_L^\varkappa(\lambda)$. Namely, for $\lambda$ 
with strictly positive imaginary part 
$$
p_L^\varkappa(\lambda)\longrightarrow p^\varkappa_\infty(\lambda), \quad
\text{as}\;\; L \rightarrow \infty.
$$
To prove this  fix a compactly supported potential. Then 
$T(\lambda)$  is an entire matrix-function, as   follows from the matrix 
exponential. Consider now the identity 
$$
T(L,-L,\lambda) =E(\lambda L)T(\lambda)E(\lambda L)
                 = \left(\matrix a(\lambda)e^{-2i\lambda L} & b^{\star}\\
          b(\lambda)  &  a^{\star}(\lambda)e^{2i\lambda L} \endmatrix \right).
$$
Taking the trace of both sides
$$
2\Delta_{L}^{\varkappa}(\lambda) = e^{i(2\lambda L +  
\psi_L^\varkappa(\lambda))} + e^{-i(2\lambda L +  \psi_L^\varkappa(\lambda))}
= e^{-2i\lambda L} a(\lambda) + a^{\star}(\lambda)e^{2i\lambda L}.
$$
Therefore, the difference 
$
 e^{-i\psi_L^\varkappa(\lambda)} - e^{-ip_L^\varkappa(\lambda)}= 
-  e^{4i \lambda L}[  e^{\psi_L^\varkappa(\lambda)}+ a^{\star}(\lambda)]
$ 
is negligible as $L\rightarrow \infty$. That gives us a    second proof 
that $\pin$ has a positive imaginary part.

The deformation  of the measure $d\mu^\varkappa$ in the infinite-volume 
or linear limit   can best be expressed by a "commutative" diagram:
$$
\CD
d\mu_L^\varkappa(\lambda)
@>L \rightarrow \infty>>
d\mu_\infty^\varkappa(\lambda)  \\
@V\text{$\varkappa^{-1}d\mu_L^\varkappa(\lambda),\, \varkappa\downarrow 0$}VV       
@VV\text{$\varkappa^{-1}d\mu_\infty^\varkappa(\lambda),\,\varkappa \downarrow 0$}V  \\
d\mu^0_L(\lambda)={\underset n\to \sum} 
\delta(\lambda- {\pi n\over L})|\hat{\psi}(n)|^2d\lambda 
    @>L \rightarrow \infty>>
d\mu^0_\infty(\lambda)= {1\over 2\pi} |\hat{\psi}(\lambda)|^2d\lambda, 
\endCD
$$
\bigskip
\noindent{where}
$$
\hat{\psi}(n) = {1\over 2L} \int_{-L}^{+L}e^{-ix\pi n/L}\psi(x)dx,\quad \quad  
\hat{\psi}(\lambda) = \int_{R^1} e^{-ix\lambda}\psi(x) dx. 
$$
The convergence of measures has to be understood in the weak$^\ast$ sense.

\subhead 8. The scattering theory \endsubhead This section is a brief 
retreat into physics. It explains on physical grounds why $\pin$ has a positive 
imaginary part. Consider two solutions 
$f_l$ and $f_r$ with prescribed asymptotics as $x\rightarrow \pm \infty$:
$$
\xalignat  3
&\phantom{ggg}   &x  \rightarrow &- \infty  & x  \rightarrow & + \infty \\
&f_l  &f_{\rightarrow} + & B(\lambda)  f_{\leftarrow} &  A(\lambda) &
f_{\rightarrow}\\
&f_r  &D(\lambda)& f_{\leftarrow} &f_{\leftarrow}+ &C(\lambda)f_{\rightarrow}
\endxalignat
$$
where
$$
f_{\leftarrow}=\left( \matrix e^{- i\lambda x}\\
                              0\endmatrix \right), \quad
f_{\rightarrow}=\left( \matrix 0\\
                    e^{i\lambda x}\endmatrix \right)
$$ are the solutions of the free equation. The solutions $f_l$, resp. $f_r$, 
describe a particle 
which approaches a potential barrier from the left, resp. right. The 
coefficients $B$, resp. $C$, 
are called {\it left}, resp. {\it right reflection coefficients}, $A=D$ is a 
{\it transmission coefficient}. 

As in    sect. 5,  we can compute a  scattering matrix $S(\lambda)$:
$$
S(\lambda)\equiv \left(\matrix A & B \\
                               C & D \endmatrix \right)=
\left(\matrix 1/a(\lambda) & -\overline{b}(\lambda)/ a(\lambda)\\
b(\lambda) / a(\lambda) &  1/a(\lambda) \endmatrix \right).
$$
Compute probability for the particle to go through or been reflected by the 
barrier:
$$
\align
\text{\bf Prob }\{ \text{particle goes through the barrier}   \} &=\int_{R^1}
|A(\lambda)|^2 |(f(\lambda)|^2d\lambda \\
\text{\bf Prob }\{\text{particle is reflected by the barrier} \} &=\int_{R^1}
|B(\lambda)|^2 |(f(\lambda)|^2d\lambda,
\endalign
$$
where $f(\lambda),\; \int|f(\lambda)|^2d\lambda =1$, is an arbitrary 
function, see \cite{LL}. Since $\text{\bf Prob}\{\cdot\}  \leq 1$ with 
necessity we obtain $|A(\lambda)|= |e^{i  \pin}|\leq 1$, for all 
$\lambda$ and therefore $\Im \pin \geq 0$. That is a third proof that 
$\pin$ has a positive imaginary part.                           

\subhead 9. Action of $H$'s flows on $b(\lambda)$ and  the angles \endsubhead
Now we return to the main topic. In the next two sections we demonstrate that 
$I(\lambda)\equiv \Im \pin$ produces on invariant manifold a torus action
with continuum number of generators.
Note that the action-angle variables for NLS equation in the scattering case 
were constructed in \cite{ZM}. We prefer  here a method of computation which 
uses the CM and is therefore different from the conventional one.

By  direct computation, we obtain
$$
b^{\bullet}(\lambda)=\{H_1,b(\lambda)\}=-i/2 \, b(\lambda).
$$ 
Another useful fact is
$$
K \nabla b(\lambda)=\lambda J\nabla b(\lambda).
$$
Then, the  inductive rule 
$$\align
\{H_{n+1},b\} &=\phantom{-} i\int \nabla H_{n+1} J\nabla b(\lambda)\\
              &=-i\int \nabla b(\lambda) J \nabla H_{n+1}\\
              &=-i\int \nabla b(\lambda) K \nabla H_n \\
              &= \phantom{-} i\int \nabla H_n K \nabla b(\lambda) \\
              &= i\; \lambda \int \nabla H_n  J \nabla b(\lambda)\\
              &= \phantom{-} \lambda \{H_{n},b\} 
\endalign
$$
implies 
$$
b^{\bullet}(\lambda)= \{H_{n+1},b\}= -i/2 \, \lambda^n  b(\lambda). 
$$
This means that $\theta(\lambda)\equiv -\,\text{phase}\; b(\lambda)$ changes 
with a uniform speed; it is an angle  on the invariant torus.

\subhead 10. Effective solution of the moment problem \endsubhead
To show that $I(\lambda)$ is canonically conjugate to 
$\theta(\lambda)$, we establish a formula for an 
effective solution of the moment problem. 
This   was introduced in \cite{V} under some restrictive assumptions. 
Here we relax the assumptions and present   an easy proof, which can be 
generalized to the continuum case.\footnote"*"{For another approach to  
effective solution of the moment problem on the finite interval see 
\cite{F, chapter 9, volume 2}.}
\proclaim{\smc The discrete case} Let $H_n$ be  a sequence of moments of some 
positive measure $\mu_k$ on $Z^1$: $$H_{n+1}={\underset k\to \sum} k^n\mu_k.  $$ If the series
$$s_k= \sum_{n=0}^{\infty} {H_{n+1}\over n!} \left[{\sin \pi (x-k) \over 
\pi (x -k)}\right]^{(n)}_{|_{x=0}}
$$ converges absolutely for all $k$, then the measure is unique and $\mu_k=s_k$ 
for all $k$.
\endproclaim
It is easy to prove the formula for  compactly supported $\mu$. Take the 
function 
$$ 
\sin \pi (x -k)\over\pi (x -k)
$$ which vanishes everywhere on $Z^1$ except $x=k$ and consider its Taylor 
expansion
$$
{\sin \pi (x -k)\over\pi (x -k)} = 
\sum_{n=0}^{\infty} {x\over n!} \left[{\sin \pi (x -  k)\over
\pi (x -k)}\right]^{(n)}_{|_{x=0}}.
$$
Integration of  both sides with respect to $\mu$ produces  the formula. To 
prove it  under the stated assumptions one has to consider the even and 
odd parts of the measure separately:
$e_n= (\mu_n+\mu_{-n})/2, \quad o_n= (\mu_n- \mu_{-n})/2$. 
Take the even part. All moments of  odd order 
vanish and the even order moments match with the corresponding moments 
of the original measure. The formula is 
true for the measure with a cut-off, which now can be removed by the bounded 
convergence theorem.  The odd part can be considered in a similar way. The 
final result is produced by taking a sum of both parts. The proof is finished.

Denote by ${\frak G}^{\sigma}$ the Gaussian density
$$
{\frak G}^{\sigma} ={1\over \sqrt{2 \pi \sigma^2}}e^{-{x^2\over 2\sigma^2}},
\quad \sigma > 0,$$
and by 
$f^\sigma=f\otimes {\frak G}^{\sigma}$ the convolution of a function or 
measure $f$ with the kernel ${\frak G}^{\sigma}$.
\proclaim{\smc The continuum case} Let $H_n$ be  a sequence of moments of some 
measure on $R^1$. If the series
$$ 
s^{\sigma}(x)= \sum_{n=0}^{\infty} {H_{n+1}\over n!} \left[ {\frak G}^{\sigma} 
(x-y)
\right]^{(n)}_{|_{y=0}}
$$
converges absolutely for all $x$ and some $\sigma > 0$, then the measure is 
unique and $s^{\sigma}(x)= \mu^{\sigma} (x)$ for all $x$.
\endproclaim
The proof is identical to the discrete case, only now one has to integrate    
the Taylor expansion of  ${\frak G}^{\sigma}(x-y)$.

Introducing  
$I^\sigma(\lambda) $ and using the preceding formula 
$$\align
\{2 I^\sigma(\lambda'), \theta(\lambda'')\}&= \sum_{n=0}^{\infty} {1\over n!}
\left[ {\frak G}^{\sigma} (\lambda'-\lambda)\right]^{(n)}_{|_{\lambda=0}}
\{2 H_{n+1},\theta(\lambda'')\} \\
&= \sum_{n=0}^{\infty} {(\lambda'')^n  \over n!}\left[ {\frak G}^{\sigma} 
(\lambda'-\lambda)\right]^{(n)}_{|_{\lambda=0}}= {\frak G}^{\sigma} 
(\lambda'-\lambda''). \endalign
$$
Passing to the limit as $\sigma \downarrow 0$, we  obtain
$$
\{I(\lambda'),\theta(\lambda'')\}=\delta(\lambda'-\lambda'').
$$
The involutive character of the $I$'s can be proved in a similar way.

%\noindent{\bf 11. The convexity theorem.}  
\proclaim {\bf 11}{\smc The Convexity Theorem.} For a potential from 
the Schwartz' class $S(R^1)$ of complex rapidly decreasing
infinitely smooth functions, the image of the momentum map is the convex cone
of all nonnegative functions from the real part of  $S(R^1)$.
\endproclaim
The part that the map is {\it into} was already explained in  section 4. 
The  {\it onto} part is more difficult. Assume that $b(\lambda)$ is from 
$S(R^1)$ and pure real for 
definiteness. Then the theorem is proved using standard arguments, 
by reconstructing the potential from the scattering matrix, see \cite{M}. 
This involves a solution of the 
Gel'fand-Levitan integral equations for  the kernels $\Gamma_{\pm}$ of the 
transformation operators: 
$$
\Gamma_+(x,t) + \Omega(x+t) + \int\limits_{x}^{\infty} \Gamma_+(x,s) \,
\Omega (s+t) \,ds= 0,
$$
where
$$
\Omega(x)=\left(\matrix
0 & \overline{\omega}(x)\\
\omega(x) & 0
\endmatrix \right),\quad \omega(x)={1\over 2\pi}\int\limits_{-\infty}^{+\infty}
r(\lambda) e^{i\lambda x} d\lambda ,\quad 
r(\lambda) ={b(\lambda)\over a(\lambda)}
$$ 
and 
$$
\Gamma_-(x,t) + \tilde\Omega(x+t) + \int\limits^{x}_{-\infty} \Gamma_-(x,s) \,
\tilde\Omega (s+t) \,ds= 0,
$$
where
$$
\tilde\Omega(x)=\left(\matrix
0 & \overline{\tilde\omega}(x)\\
\tilde\omega(x) & 0
\endmatrix \right),\quad 
\tilde\omega(x)={1\over 2\pi}\int\limits_{-\infty}^{+\infty}
\tilde r(\lambda) e^{-i\lambda x} d\lambda ,\quad
\tilde r(\lambda) =- {\overline{b}(\lambda)\over a(\lambda)}.
$$ 
The infinite smoothness and decay of the kernels $\Gamma_{\pm}$ are provided 
by the fact that  $\Omega, \tilde\Omega$ are from $S(R^1)$.

The theorem implies that the conditions in section 4 which describe the  
function $\pin$ for potentials from the class $S(R^1)$ are not only necessary, 
but also 
sufficient for  $\pin$ to be produced by  a   potential from  $S(R^1)$.

We conjecture, by analogy with the Fourier transform  
(see section 7), that for a potential $\psi$ from the Sobolev class 
$H^s, \; s \geq 0$, {\it i.e.,} 
$(1-\Delta)^{-s/2}\psi \in L^2(R^1)$, the image of the momentum map fills up  
the convex cone of nonnegative functions of the 
weighted space $L^1(R^1,\; (1+ \lambda^2)^s\, d \lambda)$. 

\subhead 12. Solution of the Problem \endsubhead The only condition on a 
finite number 
of $H_1,\cdots, \allowbreak H_N$ to correspond some invariant
manifold is 
$$\sum\limits_{0}^{\left[{N-1\over 2}\right]} H_{i+j+1} \xi_i \xi_j >0,
\quad \text{for any } \sum\limits_{0}^{\left[{N-1\over 2}\right]} \xi^2_i >0.
$$ 
In fact, it is  the 
condition for   a truncated moment problem to be definite, see \cite{A}.

A description of the invariant manifolds in $S(R^1)$
with given values of $H_1,  \cdots,\allowbreak H_N$  reduces to a construction of
the function $\pin$ which satisfies the conditions {\it i.-iii.} of section 4 
and has the prescribed asymptotic expansion
$$
\pin= -{H_1\over \lambda}- {H_2\over \lambda^{2}}- \cdots
- {H_N\over  \lambda^N}-o({1\over \lambda^N})
$$
for $ \delta \leq \arg \lambda \leq \pi- \delta$,

We will give a description of all such functions with a given value of just one
integral $H_1$ :
$$
\pin= - {H_1\over \lambda + \varphi (\lambda)}
$$
where the  arbitrary function $\varphi(\lambda)$  is such that
\roster
\item"{\it i.}" $\varphi (\lambda)=o(1)$ for   $|\lambda| \rightarrow \infty$ 
within the angle $ \delta \leq \arg \lambda \leq \pi- \delta$;
\item"{\it ii.}" $\Im \varphi(\lambda) >0 $ for $ \Im \lambda > 0$;
\item"{\it iii.}" $\Im \varphi (\lambda)/|\lambda + \varphi(\lambda)|^2$
is of the class $S(R^1)$ on the real line. 
\endroster

Indeed, according to \cite{A, section 3.4}, {\it any} function $\pin,$ such that
$ \Im \pin > 0 \quad \text{for } \Im \lambda > 0$  and with the asymptotic 
behavior
$$ 
\pin=  -{H_1\over \lambda} -o({1\over \lambda})
$$
within the angle $ \delta \leq \arg \lambda \leq \pi- \delta$ 
can be represented in such a form with an arbitrary $\varphi(\lambda)$,   
satisfying {\it i.-ii.} and {\it vice versa}. The 
proof is easy. Consider 
$$
\varphi(\lambda)=- \lambda -{H_1\over \pin},\quad\quad \text{with } \pin = {1\over \pi}\int 
{d\mu_\infty (t) \over t- \lambda},\quad \int d\mu_\infty (t) =\pi H_1.
$$
This implies  the condition {\it i.}  Now
$$
\Im \varphi(\lambda) = -\Im \lambda + \Im \lambda \int {d \mu_\infty \over |t-\lambda|^2} 
{\pi H_1\over \left| \int{d\mu_\infty\over t-\lambda}\right|^2}.
$$
By Schwartz inequality
$$
\left|\int{d\mu_\infty\over t-\lambda}\right|^2 \leq \int d\mu_\infty 
\int{d \mu_\infty \over |t-\lambda|^2} =\pi H_1 \int{d \mu_\infty \over |t-\lambda|^2},
$$
and this implies   the condition  {\it ii.} 
The proof of converse statement is  trivial. In order to get  $\pin$ which 
corresponds to some 
invariant manifold in $S(R^1)$ we need to impose some restrictions on 
$\varphi(\lambda)$. For  real $\lambda$,  
$$
\Im\pin= {H_1 \Im \varphi(\lambda)\over |\lambda + 
\varphi(\lambda)|^2}.
$$
This produces the additional condition {\it iii.} Obviously, there are plenty 
of suitable functions $\varphi$.

The case of general $N<\infty$ can
be considered in a similar way with an aid of the so-called Schur algorithm, 
see \cite{A, section 3.3}.

\bye